\documentclass{ws-ijmpcs}
\newcommand{\etal}{{\it et al.}}
\usepackage{times}
\usepackage{txfonts}
\newcommand{\dd}{\mathrm{d}}
\newcommand{\sT}{{\scriptscriptstyle T}}

\newcommand{\pT}{\bm{p}_\sT}

\newcommand{\Tr}{\text{Tr}}
\newcommand{\mc}[1]{\mathcal{#1}}

\newcommand{\qT}{\bm{q}_\sT}
\def\trimmarks{}
\def\openaccesstext{}
\def\journalname{}
\def\titlefont{\fontsize{14.4}{18}\bfseries\boldmath\selectfont\centering{}}
\def\normalsize{\fontsize{10.95}{15}\selectfont{}}
\def\footnotesize{\fontsize{10}{14}\selectfont{}}
\def\authorfont{\normalsize}
\def\rhfont{\normalsize\itshape{}}
\def\bibfont{\normalsize}

\usepackage{hyperref} 
\hypersetup{
   pdftitle={},%
   pdfauthor={},%
   pdfsubject={},%
   pdfkeywords={},%
   pdfstartview={},%
   bookmarksopen=true, breaklinks=true, debug=true, %
   colorlinks=true, linkcolor=red, citecolor=blue, urlcolor=blue
 }

\begin{document}

\markboth{J.P. Lansberg}
{Back-to-back isolated photon-quarkonium production at the LHC}

%
%

\title{Back-to-back isolated photon-quarkonium production at the LHC and the transverse-momentum-dependent  distributions of the 
gluons in the proton}

\author{J.P. Lansberg}

\address{IPNO, Universit\'e Paris-Sud, CNRS/IN2P3, F-91406, Orsay, France
\\
Jean-Philippe.Lansberg@in2p3.fr}

\maketitle
\thispagestyle{empty}

\begin{abstract}
The study of a quasi back-to-back isolated pair made of a heavy quarkonium, such as a $J/\psi$ or a $\Upsilon$,  and a photon produced in proton-proton collisions at 
the LHC, is probably the optimal way to get right away a first experimental determination of two gluon 
transverse-momentum-dependent distributions (TMDs) in an unpolarized proton, $f_1^g$ and $h_1^{\perp\,g}$, the latter giving the 
distribution of linearly polarized gluons. To substantiate this, we calculate the transverse-momentum-dependent effects that arise in the process under study and  
discuss the feasibility of their measurements.
\end{abstract}


\section{Introduction}

It is nowadays well recognized that an important class of observables at particle colliders are 
transverse\footnote{with respect to the beam axis}-momentum distributions of final state events. 
While a transverse-momentum ($q_\sT$) distribution can be described within the standard collinear 
factorization approach for large $q_\sT$, it is convenient and sometimes necessary to resort to 
an alternative factorization approach for small $q_\sT$, typically  $q_\sT\sim M_p\ll Q$, 
$M_p$ being the proton mass and $Q$ the hard scale of the process under investigation. 
This alternative approach, called Transverse-Momentum-Dependent (TMD) factorization 
(see e.g. Ref.~\refcite{TMDfact}), takes into account the transverse motion of 
partons w.r.t.\ the direction of the parent proton. 
In this picture the small final-state transverse momentum $q_\sT$ is a consequence 
of the parton transverse momenta. 
Hence, the non-perturbative distribution functions entering a TMD factorization formula 
not only depend on the collinear momentum fractions $x$, but also on 
the transverse momentum $p_\sT$.\\

The information on the transverse-momentum dependence of unpolarized and linearly polarized gluons
in an unpolarized proton is encapsulated in the following TMD correlator
$\Phi_g$, which describes the transition from a proton to a gluon, such that 
\begin{eqnarray}\label{eq:TMDcorrelator}
\Phi_g^{\mu\nu}(x,\pT,\zeta,\mu) &\equiv&
	      2\int \frac{\dd(\xi\cdot P)\, \dd^2 \xi_\sT}{(x P\cdot n)^2 (2\pi)^3}
	      e^{i ( xP + p_\sT) \cdot \xi}
	\quad  \Tr_c \Big[ \langle P| F^{n\nu}(0)\,
	      \mc{U}_{[0,\xi]}^{n[\text{--}]}\, F^{n\mu}(\xi)\, \mc{U}_{[\xi,0]}^{n[\text{--}]}
	      |P\rangle \Big]_{\xi \cdot P^\prime = 0}\nonumber\\
&=&	-\frac{1}{2x} \bigg \{g_\sT^{\mu\nu} f_1^g
	-\bigg(\frac{p_\sT^\mu p_\sT^\nu}{M_p^2}\,
	{+}\,g_\sT^{\mu\nu}\frac{\pT^2}{2M_p^2}\bigg)
	h_1^{\perp\,g} \bigg \}\,.
\end{eqnarray}
The gluon four-momentum $p$ is decomposed as $p = xP + p_\sT + p^-n$, where 
$n$ is a lightlike dimensionless vector, conjugated to the momentum $P$ of the 
parent proton, with no transverse components and satisfying the relation
$\zeta^2 = (2n{\cdot}P)^2/n^2$
(one assumes an analogous decomposition for $k$ and one takes $p^-=k^+ =0$), $p_{\sT}^2 = -\pT^2$ and  $g^{\mu\nu}_{\sT} = 
g^{\mu\nu} - P^{\mu}n^{\nu}/P\cdot n-n^{\mu}P^{\nu}/P\cdot n$. 
In addition, $F^{\mu\nu}(x)$ is the gluon field strength. 
At leading 
twist, $\Phi_g$ is parametrized\cite{Mulders:2000sh} in terms of the two gluon TMD distributions 
discussed above, $f_1^g$ and $h_1^{\perp\,g}$. 
In Eq.~(\ref{eq:TMDcorrelator}),  the gauge link 
$\mc{U}_{[0,\xi]}^{n[\text{--}]}$ is 
needed to render the matrix element gauge invariant. It runs from $0$ to $\xi$ 
via minus infinity along the $n$ direction.

A model-independent positivity bound for $h_1^{\perp g}$ was derived in Ref.~\refcite{Mulders:2000sh} and reads
\begin{equation}
\frac{\bm p_\sT^2}{2M_p^2}\,|h_1^{\perp g}(x,\bm p_\sT^2)|\le f_1^g(x,\bm p_\sT^2)\,.\label{eq:Bound}
\end{equation}
Since it is $T$-even, $h_1^{\perp\,g}$ does not necessarily vanish in absence of initial and final 
state interactions. This does not prevent it to be nonuniversal if it receives contributions from such interactions.

A number of suggestions to measure these unknown TMDs, $f_1^g$ and $h_1^{\perp\,g}$, have been discussed in
the literature. Whereas  $h_1^{\perp\,g}$ can, in principle, be extracted from the azimuthal dependence of the
dijet production in $pp$ collisions\cite{Boer:2009nc}, TMD-factorization breaking effects may be significant in this case due
to the existence of both initial and final state interactions\cite{Rogers:2010dm}. To avoid these, one may prefer to rely
on heavy-quark or dijet {\it electro}-production\cite{ep2jetLO1,ep2jetLO2} which would however only be studied at a future EIC facility.
Diphoton production\cite{Qiu:2011ai} should also not be sensitive to factorization-breaking effects. It however suffers from a huge $\pi^0$ background
and from contaminations from quark-induced channels at RHIC energies which may preclude a clean gluon TMD extraction. It has recently been proposed\cite{Boer:2012bt}
to look at $C=+1$ quarkonium production in the region where their transverse momentum is smaller than their mass. A one-loop analysis has shown
that $^1S_0$ production may be safer\cite{Ma:2014oha} than that of $^3P_J$ where color-octet contributions might spoil factorization\cite{Ma:2014oha}.
Low-$q_T$ $C=+1$ quarkonium production may however be very challenging at the LHC; a first study of $\eta_c$ production has just been carried out
by the LHCb but for $q_T> 6$ GeV. Prospects are certainly more promising with the proposed LHC fixed-target experiment AFTER@LHC(see e.g. Ref.~\refcite{Brodsky:2012vg,Lansberg:2012kf,Lansberg:2014myg,Massacrier:2015nsm}).

We claim here that polarized and unpolarized  gluon TMDs can be accessed right now at the LHC
through the reaction\cite{Dunnen:2014eta}
\begin{equation}
p(P_A)+p(P_B)\to {\cal Q}(P_{\cal Q}) + \gamma (P_{\gamma})+ X\,,
\end{equation}
 where now ${\cal Q}$ is one of the $C=-1$ charmonium or bottomonium (e.g.~$J/\psi$ or $\Upsilon$) produced almost back-to-back with a photon. In this case, the momentum imbalance of the pair in the final state, $\qT = \bm P_{{\cal Q}\sT}+\bm P_{\gamma\sT}$, is small, but not the individual transverse momenta of the two particles. These should be well detectable by
the ATLAS or CMS detectors for instance. As we discussed in Ref.~\refcite{Dunnen:2014eta}, both these final states are always produced by gluon
fusion (this is in fact also true at lower energies\cite{Lansberg:2014myg}), as illustrated by the thin and thicker curves on Fig.~\ref{CrossSec}.
In this particular configuration, the process $\Upsilon+\gamma$ is expected to be dominated by the color-singlet contributions\cite{Mathews:1999ye} 
(see the solid blue and dashed orange curves on Fig.~\ref{CrossSec}), hence TMD 
factorization should be applicable since the final-state interactions are expected to be suppressed. 
In the case of $J/\psi+\gamma$, a precise assessement of a possibe color-singlet dominance depends 
a more precise knowledge of color octet NRQCD matrix elements, which is still lacking\cite{Brambilla:2010cs,Lansberg:2008gk,Lansberg:2006dh}. 
To be on the safe side and to single out the color-singlet contribution, it could be useful to isolate the $J/\psi$.
It may also be useful to measure $J/\psi+\gamma$ production (see Ref.~\refcite{Lansberg:2009db} and references therein)
 in general in order to check the possible contribution
from Double-Parton Scatterings (DPS).

\begin{figure}[!hbt]
\centering
\includegraphics[width=0.7\columnwidth]{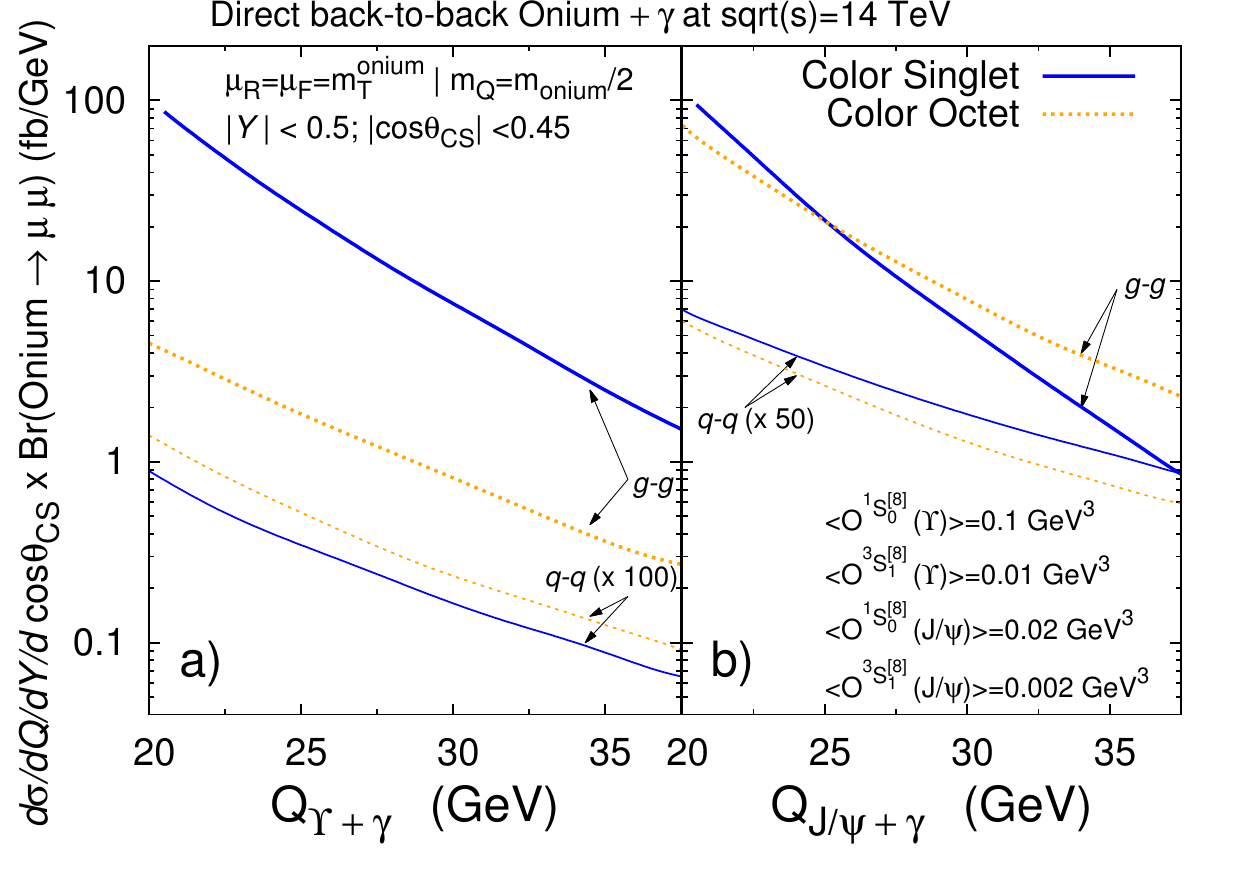}
\caption{Color-octet and color-singlet contributions from $g-g$ fusion and $q-\bar q$ annihilation channels 
 to the production of a photon with a) an $\Upsilon$ and  b) a $J/\psi$ as a function of $Q$, the invariant mass 
of the pair. The curves for the $q-\bar q$ 
annihilation are respectively rescaled by a factor 100 (50)  in a) (b)). }
\label{fig:prod_channels}
\end{figure}

It has however recently been emphasised in an analysis of Higgs plus jet production at the LHC\cite{Boer:2014lka} that it 
may not be necessary to restrict the analyses to final states made only of colorless particles for the TMD factorisation 
to be applicable, provided that one of the particles used to measure
the momentum imbalance is colorless. If this is the case, color-octet contributions to quarkonium + photon 
would not break the TMD factorisation. By extension, 
the study of the momentum imbalance in $J/\psi$ plus jet could also be used to extract gluon TMDs. At this point, 
let us mention that back-to-back $J/\psi+Z$, which has been studied by ATLAS\cite{Aad:2014kba}, could also 
be used to extract gluon TMDs. The arguments in favor of this observable are that it would probe 
them at higher scales ($Q \simeq m_Z$) and provide us with information on their evolution 
(see Ref.~\refcite{Boer:2014tka}), that in disfavor is the very small expected yield\cite{Gong:2012ah}. 
On the contrary, $J/\psi+W$ is likely contaminated by quark-induced contributions\cite{Lansberg:2013wva}.
In addition, both  $J/\psi+Z$ and  $J/\psi+W$ are likely be contaminated by significant DPS 
contributions\cite{Aad:2014rua,Aad:2014kba} ; a careful study of their suppression
by the small-momentum-imbalance requirement would  therefore be needed.
The same proviso holds for quarkonium-pair production\cite{Lansberg:2013qka,Zhang:2014vmh,Lansberg:2014swa}.


\section{TMD formalism for photon-quarkonium production}

In the TMD-factorization approach, the cross section for near back-to-back ${\cal Q}+ \gamma$ production is given by 
\begin{eqnarray}
\dd\sigma & = &\frac{1}{2 s}\,
{\int} \dd x_a \,\dd x_b \,\dd^2\bm p_{\sT,a} \,\dd^2\bm p_{\sT,b}\, \frac{\dd^3 \bm P_{\cal Q}
\dd^3 \bm P_\gamma}{16 \pi^2 E_{\cal Q} E_\gamma} \,\delta^4(p_a{+} p_b {-} P_{\cal Q} - P_\gamma)
\nonumber \\
&&\qquad \times 
{\rm Tr}\, \{ \Phi_g(x_a {,}\bm p_{\sT,a},\zeta,\mu) 
\Phi_g(x_b {,}\bm p_{ \sT,b},\zeta,\mu)
\left|{\cal A} \left (g\, g  \rightarrow {\cal Q}\,\gamma \right ) \right |^2
 \}\,,
\label{CrossSec}
\end{eqnarray}
where $s = (P_A + P_B)^2$ is the total energy squared in the hadronic centre-of-mass frame. We stress that
the $q_T$ dependence of the cross section is completely factored out from the hard-scattering amplitude 
squared $\left|{\cal A} \left (g\, g  \rightarrow {\cal Q}\,\gamma \right ) \right |^2$.

One then finds that
\begin{eqnarray}\label{eq:crosssection}\label{eq:Qgamma}
&&\frac{\dd\sigma}{\dd Q \dd Y \dd^2 \qT \dd \Omega} 
  =  \frac{4\alpha_s^2\alpha^2e^2_Q\vert R_0(0)\vert^2}{3M_{\cal Q}^2}\, \frac{Q^2-M_{\cal Q}^2 }{s Q^3 ((\gamma ^2+1)^2-(\gamma^2-1)^2 \cos^2 \theta)} \times \\
 && \times \left\{
  F_1\, \mc{C} \Big[f_1^gf_1^g\Big]
 + F_3 \,\mc{C} \Big[w_3 f_1^g h_1^{\perp g} + (x_a\! \leftrightarrow\! x_b )\Big] \cos 2\phi + \,F_4  \,\mc{C} \left[w_4 h_1^{\perp g}h_1^{\perp g}\right]\cos 4\phi  \right \}\,, \nonumber
\end{eqnarray}
where  $Q$ and $Y$ are, respectively, the invariant mass and the rapidity of the pair, $x_{a,b} =  \exp[\pm Y]\, Q/\sqrt{s}$  
and the solid angle $\Omega=(\theta,\phi)$ is measured in the Collins-Soper frame\footnote{The Collins-Soper frame is such that 
the pair is at rest and  the $\hat x\hat z$-plane spanned by 
$(\bm P_A$, $\bm P_B)$ and the $\hat x$-axis set by their bisector \cite{Collins:1977iv}.} and 
\begin{equation}
\mathcal{C}[w\, f\, g] \equiv \int \dd^{2}p_{\sT,a}  \int \dd^{2}p_{\sT,b}\,
  \delta^{2}(p_{\sT,a}+p_{\sT,b}-\bm q_{\sT})
  w(p_{\sT,a},p_{\sT,b})\, f(x_1,p_{\sT,a}^{2})\, g(x_2,p_{\sT,a}^{2})
.\nonumber
\end{equation}
In Eq.~(\ref{eq:Qgamma}), $R_0(0)$ is the quarkonium radial wave function at the origin in the position space 
and $e_Q$ is the heavy-quark charge in units of the proton 
charge. The factors $F_i$ are given by  $ F_1  =   1 + 2 \gamma ^2 + 9 \gamma ^4 + (6 \gamma ^4-2) \cos^2\theta + (\gamma ^2-1)^2 \cos^4\theta$, 
$F_3 =   4\, \gamma^2\, \sin^2\theta$ and $ F_4  =  (\gamma ^2-1)^2 \sin^4\theta $
with $\gamma \equiv Q/M_\mc{Q}$. The explicit expressions for the 
transverse weights are  
\begin{eqnarray}
 w_3 = \frac{\qT^2\bm p_{b\sT}^2 - 2 (\qT{\cdot}\bm p_{b \sT})^2}{2 M_p^2 \qT^2},
 w_4 =  2\left[\frac{\bm p_{a\sT}{\cdot}\bm p_{b\sT}}{2M_p^2} - 
		\frac{(\bm p_{a\sT}{\cdot}\qT) (\bm p_{b\sT}{\cdot}\qT)}{M_p^2\qT^2}\right]^2 -\frac{\bm p_{a\sT}^2 \bm p_{b \sT}^2 }{4 M_p^4}.
\end{eqnarray}

\begin{figure}[t!]
\centering
{\includegraphics[width=0.45\textwidth]{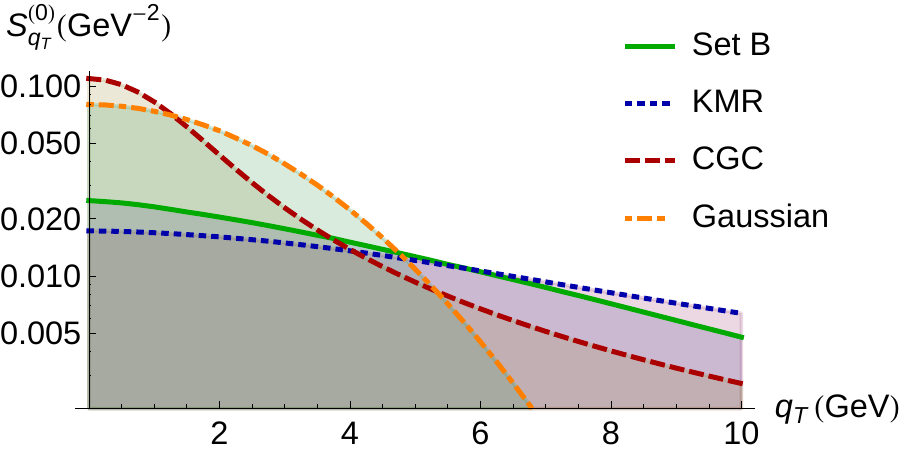}}
{\includegraphics[width=0.45\textwidth]{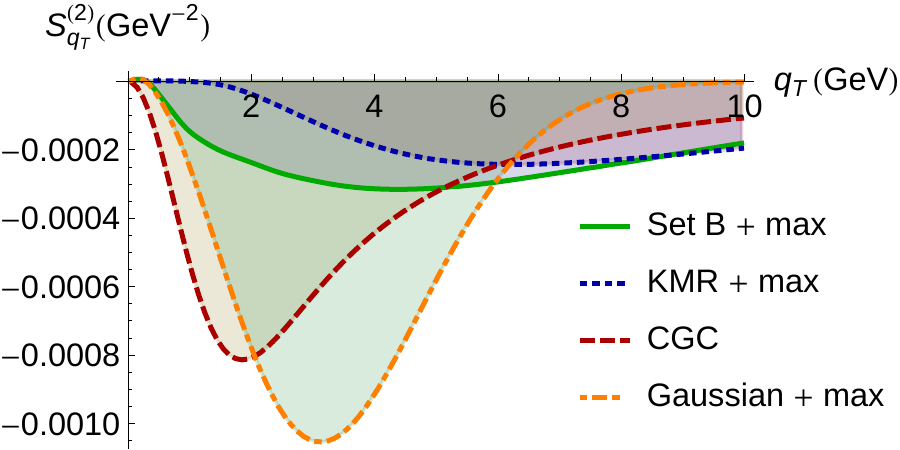}}\\
{\includegraphics[width=0.45\textwidth]{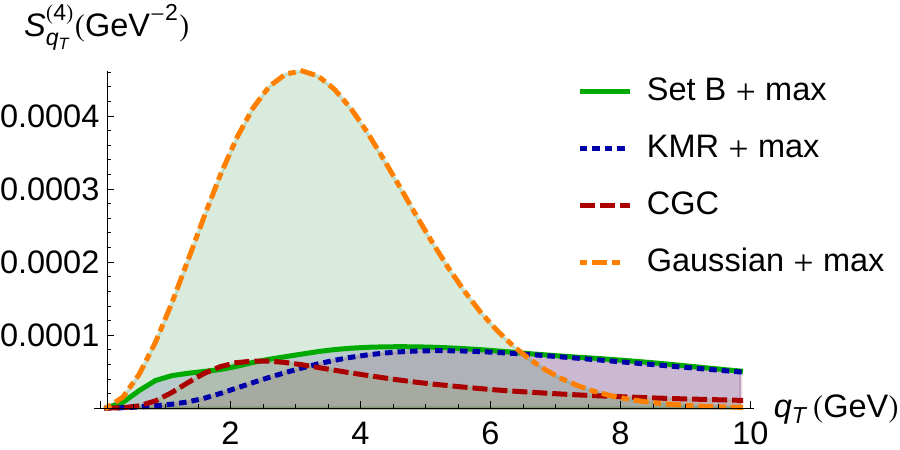}}
\caption{Evaluation of the azimutally-modulated $q_T$ spectra ${\cal S}^{(0)}_{q_T}$,  ${\cal S}^{(2)}_{q_T}$ and  ${\cal S}^{(4)}_{q_T}$ given in Eq.~(\ref{eq:qTdistrs}) for the process $p\,p\to {\cal Q}\, \gamma\,  X$, for $\sqrt{s}=14$ TeV, $Q=20$ GeV, $Y=0$, $\theta =\pi/2$.}
\label{fig:dsigma4dqT}
\end{figure}

\section{Extraction of $f_1^g$ and $ h_1^{\perp g}$}

By measuring the following $q_T$-dependent observables,
\begin{equation}
{\cal S}^{(n)}_{q_T} \equiv  \frac{\int \dd\phi\, \cos(n\, \phi )\, \frac{\dd\sigma}{\dd Q \dd Y \dd^2 \qT \dd \Omega}}
{\int_0^{Q^2/4} \dd \bm q_\sT^2 \int \dd\phi \,\frac{\dd\sigma}{\dd Q \dd Y \dd^2 \qT \dd \Omega}}\,,
\end{equation}
with  $n=0,2,4$, one can single out the three terms in Eq.~(\ref{eq:Qgamma}). We obtain
\begin{eqnarray}
\!{\cal S}^{(0)}_{q_T}\!  &=&\! \frac{\mc{C}[f_1^g f_1^g]}
  {\int  \dd \bm q_\sT^2\, \mc{C}[f_1^g f_1^g]}, \nonumber \\
  {\cal S}^{(2)}_{q_T}\!  &=&\!  
  \frac{F_3\, \mc{C}[w_3 f_1^g h_1^{\perp g} + (x_a\! \leftrightarrow\! x_b)]}
  {2 F_1 \int  \dd \bm q_\sT^2\, \mc{C}[f_1^g f_1^g]}, \nonumber \\
{\cal S}^{(4)}_{q_T}\! &=&\!  
 \frac{F_4\, \mc{C}[w_4 h_1^{\perp g} h_1^{\perp g}]}
  {2 F_1 \int \dd \bm q_\sT^2\, \mc{C}[f_1^g f_1^g]}.\label{eq:qTdistrs}
\end{eqnarray}

Fig.~\ref{fig:dsigma4dqT} shows  predictions for ${\cal S}^{(0,2,4)}_{q_T}$ for the process $\Upsilon+\gamma$
with different Ans\"atze for the TMD distributions. 
We find that the size of ${\cal S}^{(0)}_{q_T}$ should be sufficient to allow for an extraction of $f_1^g$ as a 
function of $q_\sT$. ${\cal S}^{(2)}_{q_T}$ and ${\cal S}^{(4)}_{q_T}$ are  small and one would 
need to integrate them over $\bm q_\sT^2$, (up to $Q^2/4$), to look for an experimental evidence of a 
nonzero $h_1^{\perp\,g}$.

\section{Conclusion}
 We claim that a first experimental determination of the polarized and unpolarized gluon TMD distributions 
for $x$ on the order of $10^{-3}$ can be obtained from 
the analyses of azimuthal asymmetries and transverse-momentum spectra in $p\,p \to J/\psi(\Upsilon) \, \gamma\, X$  at the LHC. 
The yields are large enough to perform these analyses using already existing data at the center-of-mass energies $\sqrt{s}=7$ and $8$ TeV.  

\section*{Acknowledgments}
I would like to thank Wilco den Dunnen, Cristian Pisano and Marc Schlegel with whom I have collaborated 
on the topic presented here. This work was supported in part by the French CNRS, 
Grants No. PICS-06149 Torino-IPNO and No. PEPS4AFTER2.

\vspace*{-.2cm}

\end{document}